\documentclass[a4paper]{article}
\RequirePackage[english]{babel}
\RequirePackage[latin1]{inputenc}
\RequirePackage[T1]{fontenc}
\RequirePackage{mathrsfs}
\RequirePackage{amsmath}
\RequirePackage{amssymb}
\RequirePackage{amsbsy}
\RequirePackage{bm}
\begin{document}
\title{\bf{On Geometrically Unified Fields and Universal Constants}}
\author{Luca Fabbri\\ 
\footnotesize INFN, Sez. di Bologna \& Dipartimento di Fisica, Universit\`{a} di Bologna\\
\footnotesize and DIPTEM Sez. Metodi e Modelli Matematici, Universit\`{a} di Genova}
\date{}
\maketitle
\begin{abstract}
We consider the Cartan extension of Riemann geometry as the basis upon which to build the Sciama--Kibble completion of Einstein gravity, developing the most general theory in which torsion and metric have two independent coupling constants: the main problem of the ESK theory was that torsion, having the Newton constant, was negligible beyond the Planck scale, but in this $\mathrm{ESK}^{2}$ theory torsion, with its own coupling constant, may be relevant much further Planck scales; further consequences of these torsionally-induced interactions will eventually be discussed.
\end{abstract}
\section*{Introduction}
When back in 1888 Gregorio Ricci-Curbastro and his pupil Tullio Levi-Civita had to found the geometry of absolute differential calculus, having no torsion was helpful because what will later be called Levi-Civita connection was symmetric and entirely written in terms of the Riemannian metric: the principle for which relativistic symmetries imposed at a differential level require the introduction of a connection could be easily implemented upon the definition of the symmetric metric Levi-Civita connection given in terms of the Riemannian metric itself.

However, some time later, the absence of torsion was recognized more as an assumption to simplify calculations than as a necessary constraint, and arguments of generality insisted for having the simplest Levi-Civita connection completed by the presence of the Cartan torsion tensor: if the principle for which relativistic symmetries imposed at a differential level require the introduction of a connection is taken in its utmost generality then the connection is not symmetric and thus torsion does not vanish, the extension of Riemann metric geometry as to include Cartan torsion being the Riemann-Cartan geometry.

The fact that there is no a priori reason to neglect torsion does not imply that there cannot be posteriori reasons implying zero torsion, and in fact when in 1916 Albert Einstein had to construct the theory of gravitation forcing torsion to vanish was insightful, because what will later be known as Einstein tensor was symmetric and divergenceless; such a physical quantity was essential since at that time physics described only systems with energy densities symmetric and divergenceless: the link between spacetime and matter could be realized by the identification of the symmetric divergenceless Einstein tensor and the symmetric divergenceless energy density known as Einstein field equations.

However, few decades after that, the vanishing torsion should not have been welcomed any longer, as there arose reasons to consider Einstein tensor no longer symmetric; the reason to need a more general geometry was that for the first time physics had a system with non-symmetric energy density and a spin density: if the demand of a link between spacetime and matter had to be extended to this more general instance then there should have been an identification of a non-symmetric Einstein tensor and the non-symmetric energy density in addition to another relationship linking torsion to the spin density in an extended system of complete Einstein-Sciama--Kibble field equations \cite{sa-si}.

In his time, Planck commented that a new scientific idea does not triumph by convincing its opponents but rather because its opponents eventually die, and a new generation grows familiar with the idea; still, torsion has not found its place beside curvature in the geometric theory of gravitation: the reason for this fact is that, if on the one hand, torsion completes the Einstein gravitation without spoiling its predictions, on the other hand, this is so because the Einstein-Sciama--Kibble gravitation as it is considered now has effects that are negligible except at the Planck scale. This makes torsion a great gift of little value.

Albeit we clearly retain that we cannot have torsional effects relevant at scales at which we know that torsional contributions have to be negligible, nevertheless if torsional effects were relevant soon beyond these scales it would be interesting to see what would follow: a torsion relevant soon after the Fermi scale means that the torsion coupling should not occur with the strength we have always been assuming, which ultimately means that in the history of torsion somewhere, somehow something about the torsion coupling constant has been overlooked; the simplest intuition may start from the remark that, because within the Einstein-Sciama--Kibble gravitational field equations there is not the identification but only the proportionality between curvature and energy and between torsion and spin, and since the curvature and torsion are independent fields, then curvature and torsion could couple to energy and spin with two different coupling constants. So far as we are concerned, the first who discussed the possibility to have two different coupling constant was Kaempffer, but despite that he clearly explained that such a modification should and could have been achieved formally, he did not discuss how to accomplish this concretely \cite{k}.

In the present paper, our purpose is to consider the Einstein-Sciama--Kibble gravitational field theory as modified by Kaempffer to implement the coupling between geometry and physics with two coupling constants, giving the explicit form of the field equations with the two different coupling constants.
\section{Kinematical Background Symmetries}
In this first section we simply recall the foundation of the Riemannian metric geometry extended as to include the Cartan torsion tensor, the RC geometry.

First, we recall the purely metric case: the Riemann geometry is based on the Riemannian metric, defining the symmetric Levi-Civita connection $\Lambda^{\alpha}_{\mu\nu}$ for the Levi-Civita metric covariant derivative $\nabla_{\mu}$ and the Riemann metric curvature tensor $R^{\alpha}_{\phantom{\alpha}\rho\nu\sigma}$ with one independent contraction $R^{\alpha}_{\phantom{\alpha}\rho\alpha\sigma}\!=\!R_{\rho\sigma}$ called Ricci metric curvature tensor itself having a single contraction $R_{\rho\sigma}g^{\rho\sigma}\!=\!R$ called Ricci metric curvature scalar; these will be obtained as the torsionless limit of more general quantities which we will next introduce. In fact if one wishes to pursue the same path Riemann followed but being in the most general case in which there is the Cartan torsion tensor, then one is taken to the Riemann-Cartan geometry, where a general connection $\Gamma^{\alpha}_{\mu\nu}$ allows to define the covariant derivative $D_{\mu}$ containing differential information, and as the most 
general connection is not symmetric
\begin{eqnarray}
&Q^{\alpha}_{\phantom{\alpha}\mu\nu}=\Gamma^{\alpha}_{\mu\nu}-\Gamma^{\alpha}_{\nu\mu}
\label{Cartan}
\end{eqnarray}
in general is not zero and it is known as Cartan torsion tensor; we assume the complete antisymmetry of Cartan torsion tensor $Q_{[\alpha\mu\rho]}\! =\!6Q_{\alpha\mu\rho}$ and the covariant constancy of the metric tensor $Dg\!=\!0$ known under the name of metricity condition, the first constraints ensuring the existence of a unique symmetric part of the connection that can be vanished in a point of a given coordinate system while the second condition ensuring that symmetric part of the connection to be vanished and the metric to be flattened in the same neighborhood of the same coordinate system. The relationship of this decomposition for the most general connection with the principle of equivalence and causality has been investigated in sets of related works, as it has been discussed in references \cite{ha,xy,a-l,m-l,f/1a,f/1b}.

From the most general connection $\Gamma^{\alpha}_{\mu\nu}$ and its derivatives we can also define
\begin{eqnarray}
&G^{\mu}_{\phantom{\mu}\rho\sigma\pi}=\partial_{\sigma}\Gamma^{\mu}_{\rho\pi}
-\partial_{\pi}\Gamma^{\mu}_{\rho\sigma}
+\Gamma^{\mu}_{\lambda\sigma}\Gamma^{\lambda}_{\rho\pi}
-\Gamma^{\mu}_{\lambda\pi}\Gamma^{\lambda}_{\rho\sigma}
\label{Riemann}
\end{eqnarray}
as Riemann curvature tensor, with one contraction $G^{\alpha}_{\phantom{\alpha}\rho\alpha\sigma}
\!=\!G_{\rho\sigma}$ that is called Ricci curvature tensor, whose only contraction $G_{\rho\sigma}g^{\rho\sigma}\!=\!G$ is called Ricci curvature scalar: both Cartan torsion and Riemann curvature tensors vanish if and only if a global coordinate system exists in which the connection vanishes.

The RC geometry can be written in world formalism by defining the dual bases of orthonormal tetrads $\xi^{a}_{\sigma}$ and $\xi_{a}^{\sigma}$ together with the introduction of the spin-connection $\Gamma^{i}_{j\mu}$ defining the covariant derivative $D_{\mu}$ that extends the differential properties to this formalism; what corresponds to the metric tensor are the Minkowskian matrices $\eta_{aq}$ and $\eta^{aq}$ as known: the previously introduced formalism of coordinate indices and the presently defined formalism of world indices are made equivalent upon the requirement of the covariant constancy of the tetrads and the Minkowskian matrices $D\xi=0$ and $D\eta=0$ which we will call formalism-compatibility conditions. Although in this formalism it is not possible to define torsion, the torsion tensor (\ref{Cartan}) can be written as
\begin{eqnarray}
&-Q^{a}_{\phantom{a}\mu\nu}=\partial_{\mu}\xi^{a}_{\nu}-\partial_{\nu}\xi^{a}_{\mu}
+\Gamma^{a}_{j\mu}\xi^{j}_{\nu}-\Gamma^{a}_{j\nu}\xi^{j}_{\mu}
\label{Cartangauge}
\end{eqnarray}
identically, as it is easy to check with a straightforward computation.

Now considering the spin-connection it is possible to define
\begin{eqnarray}
&G^{a}_{\phantom{a}b\sigma\pi}
=\partial_{\sigma}\Gamma^{a}_{b\pi}-\partial_{\pi}\Gamma^{a}_{b\sigma}
+\Gamma^{a}_{j\sigma}\Gamma^{j}_{b\pi}-\Gamma^{a}_{j\pi}\Gamma^{j}_{b\sigma}
\label{Riemanngauge}
\end{eqnarray}
which is the Riemann curvature tensor written in this formalism.

We also recall that it is possible to define a geometry of complex fields, with gauge-connection $A_{\mu}$ defining gauge-covariant derivatives $D_{\mu}$ that extend the differential properties to complex fields. Here no analogous of torsion is defined.

The analogous of the curvature is given from the gauge-connection as
\begin{eqnarray}
&F_{\mu\nu}=\partial_{\mu}A_{\nu}-\partial_{\nu}A_{\mu}
\label{Maxwellgauge}
\end{eqnarray}
called Maxwell tensor, where $q$ is called the charge of the complex field.

Writing the RC geometry in the world indices formalism has the advantage that the transformation laws are now Lorentz transformations of explicit structure that can also be written in the complex representation in which the inclusion of the Maxwell geometry would fit perfectly: this can be done after the introduction of the matrices $\boldsymbol{\gamma}_{a}$ verifying $\{\boldsymbol{\gamma}_{i},\boldsymbol{\gamma}_{j}\}=2\boldsymbol{\mathbb{I}}\eta_{ij}$ by defining the set of matrices $\boldsymbol{\sigma}_{ab}$ given by $[\boldsymbol{\gamma}_{i},\boldsymbol{\gamma}_{j}]=4\boldsymbol{\sigma}_{ij}$ defining $\{\boldsymbol{\gamma}_{i},\boldsymbol{\sigma}_{jk}\}=i\varepsilon_{ijkq}\boldsymbol{\gamma}\boldsymbol{\gamma}^{q}$ which can be proven to be a set of complex generators of the infinitesimal Lorentz complex transformation we need, called spinorial transformation, and then the spinor-connection $\boldsymbol{A}_{\mu}$ defines the spinor-covariant derivative $\boldsymbol{D}_{\mu}$ that contains the information about the dynamics of the spinor fields, of which we are going to consider the simplest $\frac{1}{2}$-spin spinor field alone; the spinorial constancy of the matrices $\boldsymbol{\gamma}_{j}$ is implemented automatically. No analogous of torsion is defined.

The analogous of the Riemann curvature given with the spinor-connection
\begin{eqnarray}
&\boldsymbol{F}_{\sigma\pi}
=\partial_{\sigma}\boldsymbol{A}_{\pi}-\partial_{\pi}\boldsymbol{A}_{\sigma}
+[\boldsymbol{A}_{\sigma},\boldsymbol{A}_{\pi}]
\label{RiemannMaxwellgauge}
\end{eqnarray}
which is a tensorial spinor antisymmetric in the tensorial indices.

As a final comment, we remark that this kinematic background has been constructed by requiring only the implementation of general symmetry principles for the underlying geometry, where once the rototraslations are gauged then the tetrad-basis and spin-connection are potentials while the Cartan and Riemann tensors are strengths according to (\ref{Cartangauge}-\ref{Riemanngauge}) of the gravitational field, as it has been demonstrated in references \cite{h,h-h-k-n}, analogously to the fact that once the phase transformation is gauged then the gauge field is potential while Maxwell tensor is strength according to (\ref{Maxwellgauge}) of the electrodynamic field.
\section{Dynamical Field Equations}
Having settled the RC kinematics, next we are going to consider the Einsteinian purely metric gravity completing it as to include the Sciama--Kibble torsional sector, obtaining the known ESK theory; then we consider the suggestion of Kaempffer to enlarge the ESK model as to have both torsion and metric entering with their own coupling constant, developing what we call the $\mathrm{ESK}^{2}$ theory.

First of all, we recall the historical path: when Einstein asked torsion to vanish he was motivated by the fact that in doing so the Bianchi identities in their contracted form $\nabla_{\mu}\!\!\left(R^{\mu\nu}\!-\!\frac{1}{2}g^{\mu\nu}R\!-\!\lambda g^{\mu\nu}\right)\!\equiv\!0$ suggested that the so called Einstein tensor $R_{\mu\nu}\!-\!\frac{1}{2}g_{\mu\nu}R$ would be symmetric and divergenceless like all energy density tensors $T_{\mu\nu}$ known at that time; in searching for field equations linking geometrical quantities on the one hand and material fields on the other, he thought to set $R_{\mu\nu}-\frac{1}{2}g_{\mu\nu}R-\lambda g_{\mu\nu}=8\pi kT_{\mu\nu}$ in terms of some proportionality constant $k$ later acknowledged to be the Newton constant. These field equations might have been obtained through the variation of the lagrangian density that is given in terms of the Ricci metric curvature scalar alone, which is the only possible torsionless least-order derivative lagrangian density. Now if one wishes to pursue the same path that Einstein followed but being in the most general case in which there is the torsion tensor, then the torsional Bianchi identities, which we call Jacobi-Bianchi identities to distinguish them, can be obtained, and when they are written in their fully contracted form it is straightforward to see that they can be worked out to be given by the geometrical identities
\begin{eqnarray}
&D_{\rho}Q^{\rho\mu\nu}\!-\!\left(G^{\mu\nu}-\frac{1}{2}g^{\mu\nu}G-\lambda g^{\mu\nu}\right)
\!+\!\left(G^{\nu\mu}-\frac{1}{2}g^{\nu\mu}G-\lambda g^{\nu\mu}\right)\equiv0\\
&D_{\mu}\!\!\left(G^{\mu\nu}\!-\!\frac{1}{2}g^{\mu\nu}G\!-\!\lambda g^{\mu\nu}\right)
\!\!-\!\!\left(G_{\rho\beta}\!-\!\frac{1}{2}g_{\rho\beta}G\!
-\!\lambda g_{\rho\beta}\right)\!Q^{\beta\rho\nu}
\!\!+\!\frac{1}{2}G^{\mu\rho\beta\nu}Q_{\beta\mu\rho}\!\equiv\!0
\end{eqnarray}
and one should also take into account the fact that an additional spin density tensor $S_{\lambda\mu\nu}$ is present beside an energy density tensor $T_{\mu\nu}$ that is not symmetric any longer; in searching for field equations linking geometrical quantities on the one hand and material fields on the other, we may still define Einstein tensor in the same way and eventually set the field equations given by 
\begin{eqnarray}
&Q^{\rho\mu\nu}=-16\pi kS^{\rho\mu\nu}\\
&G^{\mu}_{\phantom{\mu}\nu}-\frac{1}{2}\delta^{\mu}_{\nu}G
-\lambda\delta^{\mu}_{\nu}=8\pi kT^{\mu}_{\phantom{\mu}\nu}
\end{eqnarray}
so to convert the above identities into the conservation laws
\begin{eqnarray}
&D_{\rho}S^{\rho\mu\nu}+\frac{1}{2}\left(T^{\mu\nu}-T^{\nu\mu}\right)\equiv0\\
&D_{\mu}T^{\mu\nu}+T_{\rho\beta}Q^{\rho\beta\nu}-S_{\mu\rho\beta}G^{\mu\rho\beta\nu}\equiv0
\end{eqnarray}
which are to be valid once the matter field equations are assigned. These field equations may be obtained through the variation of the lagrangian density given in terms of the Ricci scalar $\mathscr{L}\!=\!\frac{1}{16\pi k}(G\!+\!2\lambda)$ being it the simplest torsional completion of the least-order derivative lagrangian density. However, although this system of field equations has the Einstein equations completed to include torsion, nevertheless this is only the most straightforward but not yet the most general of all the possible enlargements in which torsion is present, with a spin density tensor $S_{\lambda\mu\nu}$ and energy density tensor $T_{\mu\nu}$ not symmetric; in searching for field equations linking geometrical quantities on the one hand and material fields on the other, we have that the field equations given by the following
\begin{eqnarray}
&Q^{\rho\mu\nu}=-aS^{\rho\mu\nu}\\
&\frac{b}{2a}\!\left(\frac{1}{4}\delta^{\mu}_{\nu}Q^{2}
\!-\!\frac{1}{2}Q^{\mu\alpha\sigma}Q_{\nu\alpha\sigma}
\!+\!D_{\rho}Q^{\rho\mu}_{\phantom{\rho\mu}\nu}\right)
\!+\!\left(G^{\mu}_{\phantom{\mu}\nu}\!-\!\frac{1}{2}\delta^{\mu}_{\nu}G -\lambda\delta^{\mu}_{\nu}\right)\!=\!\left(\frac{b+a}{2}\right)T^{\mu}_{\phantom{\mu}\nu}
\end{eqnarray}
are in fact the most general for which the above Jacobi-Bianchi identities can be used to obtain the validity of the above conservation laws
\begin{eqnarray}
&D_{\rho}S^{\rho\mu\nu}+\frac{1}{2}\left(T^{\mu\nu}-T^{\nu\mu}\right)\equiv0\\
&D_{\mu}T^{\mu\nu}+T_{\rho\beta}Q^{\rho\beta\nu}-S_{\mu\rho\beta}G^{\mu\rho\beta\nu}\equiv0
\end{eqnarray}
holding when the matter field equations are assigned. To see that these are the most general field equations we just have to notice that they may be obtained through the variation of the lagrangian density given in terms of the Ricci scalar plus torsional squared contributions, so that torsion is both implicitly and explicitly present beside the curvature scalar, in a lagrangian density given by the form $\mathscr{L}\!=\!-\frac{1}{4a}Q^{2}\!+\!\frac{1}{(a+b)}(R\!+\!2\lambda)\!\equiv\!
-\frac{b}{4a(a+b)}Q^{2}\!+\!\frac{1}{(a+b)}(G\!+\!2\lambda)$ which is the most general completely antisymmetric torsion completion of the least-order derivative lagrangian density as it can be proven quite straightforwardly.

Next step would then consist in the inclusion of the Maxwell field, for which we may follow a path that is analogous to the previous one by having some geometrical identities in the case of gauge fields to suggest the form that the gauge field equation should have, and in order to do so we can use the commutator of the gauge curvature taken in its fully contracted form as
\begin{eqnarray}
&D_{\rho}\left(D_{\sigma}F^{\sigma\rho}+\frac{1}{2}F_{\alpha\mu}Q^{\alpha\mu\rho}\right)\equiv0
\end{eqnarray}
and considering that also the current vector $J_{\mu}$ is to be accounted; for field equations linking geometrical quantities and material fields, we have the field equations for the completely antisymmetric torsion and curvature tensors as
\begin{eqnarray}
&Q^{\rho\mu\nu}=-aS^{\rho\mu\nu}
\label{torsion-spin}\\
\nonumber
&\frac{b}{2a}\left(\frac{1}{4}\delta^{\mu}_{\nu}Q^{2}
-\frac{1}{2}Q^{\mu\alpha\sigma}Q_{\nu\alpha\sigma}
+D_{\rho}Q^{\rho\mu}_{\phantom{\rho\mu}\nu}\right)
+\left(G^{\mu}_{\phantom{\mu}\nu}-\frac{1}{2}\delta^{\mu}_{\nu}G
-\lambda\delta^{\mu}_{\nu}\right)+\\
&+\left(\frac{b+a}{2}\right)\left(F^{\rho\mu}F_{\rho\nu}-\frac{1}{4}\delta^{\mu}_{\nu}F^{2}\right)
=\left(\frac{b+a}{2}\right)T^{\mu}_{\phantom{\mu}\nu}
\label{curvature-energy}
\end{eqnarray}
together with the field equations for the gauge fields as
\begin{eqnarray}
&\frac{1}{2}F_{\alpha\mu}Q^{\alpha\mu\rho}+D_{\sigma}F^{\sigma\rho}=J^{\rho}
\label{gauge-current}
\end{eqnarray}
so to convert the above geometrical identities into the conservation laws for the completely antisymmetric spin and energy densities given by
\begin{eqnarray}
&D_{\rho}S^{\rho\mu\nu}+\frac{1}{2}\left(T^{\mu\nu}-T^{\nu\mu}\right)\equiv0
\label{conservationspin}\\
&D_{\mu}T^{\mu\nu}
+T_{\rho\beta}Q^{\rho\beta\nu}-S_{\mu\rho\beta}G^{\mu\rho\beta\nu}+J_{\rho}F^{\rho\nu}\equiv0
\label{conservationenergy}
\end{eqnarray}
and also for the current given by the expression
\begin{eqnarray}
&D_{\rho}J^{\rho}=0
\label{conservationcurrent}
\end{eqnarray}
which are valid as the matter field equations are assigned. These field equations can also be obtained by varying the previous gravitational lagrangian density plus the electrodynamic lagrangian density $\mathscr{L}\!=\!-\frac{1}{4}F^{2}$ as the most comprehensive geometric least-order derivative lagrangian density as it is rather clear.

This is the geometrical system of field equations defining the structure of the torsional-gravitational gauge interactions which has to be coupled to a given material content, and the completely antisymmetric spin and energy densities 
\begin{eqnarray}
&S^{\rho\mu\nu}
=\frac{i\hbar}{4}\overline{\psi}\{\boldsymbol{\gamma}^{\rho},\boldsymbol{\sigma}^{\mu\nu}\}\psi
\label{spin}\\
&T^{\mu}_{\phantom{\mu}\nu}
=\frac{i\hbar}{2}\left(\overline{\psi}\boldsymbol{\gamma}^{\mu}\boldsymbol{D}_{\nu}\psi
-\boldsymbol{D}_{\nu}\overline{\psi}\boldsymbol{\gamma}^{\mu}\psi\right)
\label{energy}
\end{eqnarray}
alongside with the current given by the expression
\begin{eqnarray}
&J^{\rho}=q\hbar\overline{\psi}\boldsymbol{\gamma}^{\rho}\psi
\label{current}
\end{eqnarray}
are precisely the conserved quantities we need once the spinor field has
\begin{eqnarray}
&i\hbar\boldsymbol{\gamma}^{\mu}\boldsymbol{D}_{\mu}\psi-m\psi=0
\label{matterequations}
\end{eqnarray}
as the matter field equations themselves. These field equations might have also been obtained by varying the previous geometric lagrangian density plus Dirac lagrangian density $\mathscr{L}\!=\!
\frac{i\hbar}{2}(\overline{\psi}\boldsymbol{\gamma}^{\mu}\boldsymbol{D}_{\mu}\psi\!-\!
\boldsymbol{D}_{\mu}\overline{\psi}\boldsymbol{\gamma}^{\mu}\psi)\!-\!m\overline{\psi}\psi$ as the entire least-order derivative lagrangian density that can be written for this theory as a whole. 

Taken all together, we have that the most general coupling involving the completely antisymmetric torsion present both implicitly through the connection and explicitly with quadratic terms beside the curvature scalar, plus gauge fields, for the Dirac matter content, is given by the system of field equations for the completely antisymmetric torsion-spin coupling (\ref{torsion-spin}-\ref{spin}) together with curvature-energy coupling given in terms of (\ref{curvature-energy}-\ref{energy}) and alongside to gauge-current coupling given by (\ref{gauge-current}-\ref{current}) given by the geometric field equations for the completely antisymmetric torsion and curvature tensor with the gauge field
\begin{eqnarray}
&Q^{\rho\mu\nu}
=-a\frac{i\hbar}{4}\overline{\psi}\{\boldsymbol{\gamma}^{\rho},\boldsymbol{\sigma}^{\mu\nu}\}\psi
\label{Sciama--Kibble}\\
\nonumber
&\frac{b}{2a}\left(\frac{1}{4}\delta^{\mu}_{\nu}Q^{2}
-\frac{1}{2}Q^{\mu\alpha\sigma}Q_{\nu\alpha\sigma}
+D_{\rho}Q^{\rho\mu}_{\phantom{\rho\mu}\nu}\right)
+\left(G^{\mu}_{\phantom{\mu}\nu}-\frac{1}{2}\delta^{\mu}_{\nu}G
-\lambda\delta^{\mu}_{\nu}\right)+\\
&+\left(\frac{b+a}{2}\right)\left(F^{\rho\mu}F_{\rho\nu}-\frac{1}{4}\delta^{\mu}_{\nu}F^{2}\right)
=\left(\frac{b+a}{2}\right)
\frac{i\hbar}{2}\left(\overline{\psi}\boldsymbol{\gamma}^{\mu}\boldsymbol{D}_{\nu}\psi
-\boldsymbol{D}_{\nu}\overline{\psi}\boldsymbol{\gamma}^{\mu}\psi\right)
\label{Einstein}\\
&\frac{1}{2}F_{\mu\nu}Q^{\mu\nu\rho}+D_{\sigma}F^{\sigma\rho}
=q\hbar\overline{\psi}\boldsymbol{\gamma}^{\rho}\psi
\label{Maxwell}
\end{eqnarray}
verified once the matter field equations (\ref{matterequations}) given by
\begin{eqnarray}
&i\hbar\boldsymbol{\gamma}^{\mu}\boldsymbol{D}_{\mu}\psi-m\psi=0
\label{Dirac}
\end{eqnarray}
are satisfied, as a direct calculation shows: notice that a completely antisymmetric torsion restrains the description to a completely antisymmetric spin allowing only the simplest spinor field to be defined without constraints, or conversely requiring complete antisymmetry for torsion does not constitute any loss of generality for the complete antisymmetry of the spin since we want to study only the simplest spinor field; then the matter field equation (\ref{Dirac}) has characteristic equation given simply by $n^{2}\!=\!0$ so that $n^{\mu}$ is light-like thus yielding the characteristic surfaces on the light-cone and causality is preserved. The fact that all degrees of freedom are accounted by a causal matter field equation tells that the matter field equation is well defined, as it has been discussed in \cite{f/2a,f/2b}.

As a final remark, we notice that these dynamical fields have been endowed with equations for the completely antisymmetric torsion-spin and curvature-energy coupling with gauge-current coupling for a geometry filled with Dirac matter fields, where the completely antisymmetric torsion and gravitational constants $a$ and $b$ are accompanied by the electric charge $q$ and the Planck constant $\hbar$ altogether accounting for as many free coupling constants as independent fields, since the cosmological constant $\lambda$ and the mass $m$ of the spinor have to be seen as parameters; again we stress that if we wish to develop a geometry of Dirac matter fields whose completely antisymmetric spin turns into a completely antisymmetric torsion for least-order derivative field equations then this theory is the only one possible, but if one releases the assumption of working with Dirac matter fields then there no longer is a complete antisymmetry of the spin that is to be reflected into the complete antisymmetry of torsion and if this is accompanied by a further releasing of the hypothesis of having only the least-order derivative field equations, then more terms quadratic in the curvature and quartic in torsion can be added in enlarged actions such as those for instance of reference \cite{Baekler:2011jt}, or with even more curvature and torsion in yet even more enlarged actions, including more and more coupling constants.

However, in order to see what happens to the Dirac field with completely antisymmetric spin coupled to the completely antisymmetric torsion in least-order derivative field equations, our restrictions will work just fine.
\subsection{The Self-Interactions for Matter Fields}
In the system of field equations, torsional quantities can be decomposed in terms of torsionless quantities and torsional contributions that can be converted through the torsion-spin coupling field equation into spinorial potentials in all field equations starting from the curvature-energy coupling field equations as
\begin{eqnarray}
\nonumber
&\left(R_{\mu\nu}+\lambda g_{\mu\nu}\right)
+\left(\frac{a+b}{2}\right)\left(F^{\rho}_{\phantom{\rho}\mu}F_{\rho\nu}-\frac{1}{4}g_{\mu\nu}F^{2}\right)
=-\left(\frac{a+b}{2}\right)\frac{m}{2}\overline{\psi}\psi g_{\mu\nu}+\\
&+\left(\frac{a+b}{2}\right)
\frac{i\hbar}{4}\left(\overline{\psi}\boldsymbol{\gamma}_{\mu}\boldsymbol{\nabla}_{\nu}\psi
+\overline{\psi}\boldsymbol{\gamma}_{\nu}\boldsymbol{\nabla}_{\mu}\psi
-\boldsymbol{\nabla}_{\nu}\overline{\psi}\boldsymbol{\gamma}_{\mu}\psi
-\boldsymbol{\nabla}_{\mu}\overline{\psi}\boldsymbol{\gamma}_{\nu}\psi\right)
\label{gravitation}
\end{eqnarray}
which are exactly the field equations we would have had without torsion and for the gauge-current coupling field equations given by
\begin{eqnarray}
&\nabla_{\sigma}F^{\sigma\rho}=q\hbar\overline{\psi}\boldsymbol{\gamma}^{\rho}\psi
\label{electrodynamics}
\end{eqnarray}
again as those we would have had with no torsion, with matter field equations
\begin{eqnarray}
\nonumber
&i\hbar\boldsymbol{\gamma}^{\mu}\boldsymbol{\nabla}_{\mu}\psi+\frac{3a}{16} \hbar^{2}\overline{\psi}\boldsymbol{\gamma}^{\mu}\boldsymbol{\gamma}\psi
\boldsymbol{\gamma}_{\mu}\boldsymbol{\gamma}\psi-m\psi\equiv\\
\nonumber
&\equiv i\hbar\boldsymbol{\gamma}^{\mu}\boldsymbol{\nabla}_{\mu}\psi-\frac{3a}{16} 
\hbar^{2}\overline{\psi}\boldsymbol{\gamma}^{\mu}\psi\boldsymbol{\gamma}_{\mu}\psi-m\psi\equiv\\
&\equiv i\hbar\boldsymbol{\gamma}^{\mu}\boldsymbol{\nabla}_{\mu}\psi
-\frac{3a}{16}\hbar^{2}\left(\overline{\psi}\psi\mathbb{I}
-\overline{\psi}\boldsymbol{\gamma}\psi\boldsymbol{\gamma}\right)\psi-m\psi=0
\label{matter}
\end{eqnarray}
as in the torsionless case but complemented with self-interactions for spinors having the Nambu-Jona--Lasinio structure, and therefore the torsionally-induced self-interactions for a given spinor are actually chiral interactions between the projections of each spinor, within the Dirac field equations.

In non-gravitational non-relativistic limit the temporal and spatial components are split, so that for electric and magnetic fields and taking the standard representation where the spinor field has only the large component $\phi$ then
\begin{eqnarray}
&\mathrm{div}\vec{E}=q\hbar\phi^{\dagger}\phi\\
&\mathrm{rot}\vec{B}-\frac{\partial\vec{E}}{\partial t}=0
\end{eqnarray}
with matter field equations given according to
\begin{eqnarray}
\nonumber
&i\hbar\frac{\partial\phi}{\partial t}
+\frac{\hbar^{2}}{2m}\boldsymbol{\nabla}^{2}\phi
-\frac{q\hbar^{2}}{2m}\vec{B}\cdot\vec{\boldsymbol{\sigma}}\phi
-\frac{3a}{16}\hbar^{2}\phi^{\dagger}\vec{\boldsymbol{\sigma}}\phi
\cdot \vec{\boldsymbol{\sigma}}\phi-m\phi\equiv\\
&\equiv i\hbar\frac{\partial\phi}{\partial t}
+\frac{\hbar^{2}}{2m}\boldsymbol{\nabla}^{2}\phi
-\frac{q\hbar^{2}}{2m}\vec{B}\cdot\vec{\boldsymbol{\sigma}}\phi
-\frac{3a}{16}\hbar^{2}\phi^{\dagger}\phi\phi-m\phi=0
\label{matterapproximated}
\end{eqnarray}
where the presence of torsion is manifested as semispinorial self-interactions of the Ginzburg-Landau form, and these are known as Pauli field equations.

As a final remark, we have to notice that when we consider a static configurations of total energy $E$ it is possible to have situations where the semispinor has a single component $\phi^{\dagger}\!=\!(u^{\ast},0)$ or $\phi^{\dagger}\!=\!(0,u^{\ast})$ with matter field equation
\begin{eqnarray}
&\frac{\hbar^{2}}{2m}\boldsymbol{\nabla}^{2}u-\frac{3a}{16}\hbar^{2}u^{\ast}u u+Eu=0
\label{matterapproximatedlimit}
\end{eqnarray}
in which the torsion is eventually expressed in the guise of semispinorial self-interactions with Gross-Pitaevskii form, in the Schr\"{o}dinger equation.

We have to notice that the constant $a\!+\!b\!=\!16\pi G_{N}$ is to be interpreted as the gravitational Newton constant, the constant $q$ is of course the electric charge and the constant $\frac{3a\hbar^{2}}{16}$ is still totally undetermined: in the case in which the torsional constant is taken to be positive therefore giving rise to a repulsion in the non-linear potentials, since for antialigned-spin matter distributions the overall non-linear potential vanishes, the non-linear potential that keeps apart two matter distributions in general fails to do so by allowing linear superposition in the case of opposite helicities, consequently entailing a dynamical form of the exclusion principle, as it has been discussed in \cite{f/3}: in particular, as for single-handed massless fields the non-linear potential vanishes, neutrinos would not obey the exclusion principle, as suggested in \cite{d-s}; the fact that the torsional constant is positive also implies that at high densities the dominant forces are repulsive, whereas the fact that the total energy can also be negative tells that at low densities the dominant effect is attractiveness, and hence the whole potential is capable of giving rise to a dynamical symmetry breaking down to a stable equilibrium with a positive energy: condition $3a\hbar^{2}\phi^{2}\!=\!16E$ is the non-trivial solution we have for bosonization and the eventual condensation in the theory of superconductivity as it is well known. Therefore we may say that this approach is what gives rise to a geometric justification of superconductivity theories.

With what we have done so far, we have presented a theory that allows for the possibility to have the torsional interactions and coupling constant turned into spinorial non-linear potentials with a totally undetermined strength, a theory in which the non-linearities would become more relevant as the strength gets larger, within the Dirac matter field equation: in this theory the non-linear effects may be amplified by a larger strength rendering them more likely to be detected in the sense that in the restricted theory such non-linearities are tuned on the gravitational constant remaining beyond observation, although the fact that non-linear effects may be amplified by a larger strength does not imply this will necessarily be the case, for the dynamic of the Dirac matter fields.

Because having torsion with a constant that is left undetermined would always allow for the possibility to set it to any specific value experimentally while fixing its constant immediately would forbid any eventual tuning, the wisest choice is to take torsion with its constant undetermined.

And therefore we will have to let observation tell.
\section*{Conclusion}
In the present paper, we have started from the ESK theory considering Kaempffer speculation for which it must have been possible to incorporate for the independent metric and torsion fields two different coupling constants, giving to this $\mathrm{ESK}^{2}$ theory a concrete realization; we have obtained all the field equations, which have eventually been decomposed and rearranged in order to show that the presence of torsion and its coupling constant are converted into spinorial self-interactions displaying a free universal coupling constant within the Dirac matter field equations: we have shown that the non-linear potentials have the structure of the Nambu-Jona--Lasinio potential in the Dirac spinorial equation, approximated to the structure of the Ginzburg-Landau potential in the Pauli non-relativistic semi-spinorial equation, and then to the structure of the Gross-Pitaevskii potential in the Schr\"{o}dinger non-relativistic scalar--like equation.

The main result of the present work is that the torsional spinorial self-interactions may give rise to the potentials related to a constant $a$ that in the ESK theory is the Newton constant, whose smallness suppresses all sorts of effects unless at the Planck scale, whereas in the present $\mathrm{ESK}^{2}$ theory the constant is still undetermined and if it is chosen properly, those effects may actually be relevant much beyond the Planck scales; this means that the single remaining argument against torsion spelling its alleged smallness does not apply any longer, and torsionally-induced interactions potentially relevant at larger scales have now the right to be studied and their consequences investigated.

Whether torsion is indeed relevant because its constant is found to be large or to prove that despite all torsion is in fact negligible for its constant happens to be small nobody will known, until detection.

\end{document}